\documentclass[twocolumn,twocolappendix]{aastex631}
\usepackage{amsmath}
\usepackage{bm}
\newcommand{\GG}[1]{}

\renewcommand{\vec}[1]{\boldsymbol{#1}}

\shorttitle{AASTeX v6.3.1 Sample article}
\shortauthors{Zhu et al.}

\graphicspath{{./}{figures/}}
\begin{document}

\title{Regulation of Proton-$\alpha$ Differential Flow by Compressive Fluctuations and Ion-scale Instabilities in the Solar Wind}

\author[0000-0002-1541-6397]{Xingyu Zhu}
\affiliation{School of Earth and Space Sciences, Peking University, Beijing 100871, People's Republic of China}
\affiliation{Mullard Space Science Laboratory, University College London, Dorking RH5 6NT, UK}
\author[0000-0002-0497-1096]{Daniel Verscharen}
\affiliation{Mullard Space Science Laboratory, University College London, Dorking RH5 6NT, UK}
\author[0000-0001-8179-417X]{Jiansen He}
\affiliation{School of Earth and Space Sciences, Peking University, Beijing 100871, People's Republic of China}
\author{Bennett A. Maruca}
\affiliation{Bartol Research Institute and Department of Physics and Astronomy, University of Delaware, Newark, DE 19716, USA}
\author[0000-0002-5982-4667]{Christopher J. Owen}
\affiliation{Mullard Space Science Laboratory, University College London, Dorking RH5 6NT, UK}

\begin{abstract}
Large-scale compressive slow-mode-like fluctuations can cause variations in the density, temperature, and magnetic-field magnitude in the solar wind. In addition, they also lead to fluctuations in the differential flow $U_{\rm p\alpha}$ between $\alpha$-particles and protons ($p$), which is a common source of free energy for the driving of ion-scale instabilities. If the amplitude of the compressive fluctuations is sufficiently large, the fluctuating $U_{\rm p\alpha}$ intermittently drives the plasma across the instability threshold, leading to the excitation of ion-scale instabilities and thus the growth of corresponding ion-scale waves. The unstable waves scatter particles and reduce the average value of $U_{\rm p\alpha}$. We propose that this ``fluctuating-beam effect''  maintains the average value of $U_{\rm p\alpha}$ well below the marginal instability threshold. We model the large-scale compressive fluctuations in the solar wind as long-wavelength slow-mode waves using a  multi-fluid model. We numerically quantify  the fluctuating-beam effect for the  Alfv\'en/ion-cyclotron (A/IC) and fast-magnetosonic/whistler (FM/W) instabilities. We show that measurements of the proton-$\alpha$ differential flow and compressive fluctuations from the {\it Wind} spacecraft are consistent with our predictions for the fluctuating-beam effect. This effect creates a new channel for a direct cross-scale energy transfer from large-scale compressions to ion-scale fluctuations.
\end{abstract}

\section{Introduction} \label{sec:intro}
The solar wind plasma consists of ions and electrons that are continuously energized in the solar corona and expand into the heliosphere. The plasma conditions in this environment are weakly collisional. Therefore, the velocity distribution functions of the ions often develop and maintain non-thermal features. If these features are sufficiently strong, they  provide  free energy to drive kinetic instabilities \citep{Hellinger2006, Maruca2012, Verscharen2013a, Zhao2018, Bowen2020}. 

Among the ion species in the solar wind, the $\alpha$-particles are the second-most abundant species after the protons. They typically account for about 1$-$5\% of the total ion number density and thus about 4$-$20\% of the total ion mass density \citep{Robbins1970,Bame1977}. The $\alpha$-particle abundance shows, on average, a positive correlation with the solar-wind speed \citep{Aellig2001,Kasper2007, Alterman2019}.

Protons and $\alpha$-particles often exhibit distinct bulk velocities $\vec U_{j}$, where $j=\mathrm p, \alpha$, respectively. We define the differential flow velocity between $\alpha$-particles and protons as $\vec{U}_{\mathrm p\alpha}=\vec{U}_{\rm\alpha}-\vec{U}_{\rm p}$. The resulting vector is generally parallel or anti-parallel to the local magnetic field \citep{Kasper2006} and exhibits correlations with ${U}_{\rm p}$, heliocentric distance $R$, and the collisional age \citep{Neugebauer1996,Stansby2019,Durovcova2019,Mostafavi2022}. 
Kinetic instabilities driven by proton-$\alpha$ differential flow have typical thresholds of order the local Alfv\'en speed
\begin{equation}
    V_{\rm A}\equiv\frac{B_0}{\sqrt{4\pi n_p m_p}}
\end{equation}
\citep{Verscharen2013b}, where $\vec{B}_0$ is the background magnetic field, $n_p$ is the proton number density, and $m_p$ is the proton mass. As the solar wind travels away from the Sun, $V_{\rm A}$ decreases as long as $B_0^2$ decreases faster with distance than $n_p$, which is generally true in the inner heliosphere. If we assume that all relevant plasma variables follow monotonic radial profiles and that $\vec{U}_{\mathrm p}$ and $\vec{U}_{\alpha}$ follow a ballistic trajectory, this decrease in the Alfv\'en speed with $R$ leads to an increase in $U_{\rm p\alpha}/V_{\rm A}$ with $R$, bringing $U_{\mathrm p\alpha}$ closer to the local thresholds of the instabilities driven by proton-$\alpha$ differential flow. Once $U_{\mathrm p\alpha}$ crosses a local instability threshold, small-scale fluctuations in the electromagnetic field grow at the expense of the $U_{\mathrm p\alpha}$, and, in turn, the instabilities regulate the value of $U_{\rm p\alpha}$ by limiting it to the local threshold \citep{Marsch1982,Marsch1987,Gary2000a,Verscharen2013a}. 

The proton-$\alpha$ differential flow primarily excites two kinds of kinetic instabilities: a group of A/IC instabilities and the FM/W instability \citep{Gary1993, Verscharen2013b, Gary2016}. The parallel FM/W instability has lower thresholds than other $\alpha$-particle-driven instabilities when $\beta_{\rm p}$ is large, while the oblique A/IC instabilities have lower thresholds at large beam density or small $\beta_{\rm p}$ \citep{Gary2000a}, where 
\begin{equation}
\beta_{j}\equiv\frac{8\pi n_jk_{\mathrm B}T_{j}}{B_0^2}
\end{equation}
is the ratio of thermal energy of species $j$ to magnetic field energy, $n_j$ is the density of species $j$, $k_{\mathrm B}$ is the Boltzmann constant, and $T_j$ is the temperature of species $j$. The FM/W instability threshold strongly depends on  $T_{\rm\perp\alpha}/T_{\rm\parallel\alpha}$, $\beta_{\rm \parallel p}$ as well as $T_{\rm\parallel\alpha}/T_{\rm\parallel p}$ \citep{Gary2000b}, where $T_{\perp j}$ is the temperature of species $j$ perpendicular to the magnetic field, $T_{\parallel j}$ is the temperature of species $j$ parallel to the magnetic field, and $\beta_{\parallel j}\equiv\beta_{j}T_{\parallel j}/T_j$. \citet{Verscharen2013a} find a new, parallel A/IC instability in the parameter range $1<\beta_{\rm p}<12$. The minimum $U_{\rm p\alpha}$ for driving this parallel A/IC instability ranges from $0.7V_{\rm A}$ when $T_{\perp\alpha}=T_{\parallel \alpha}$ to $0.9V_{\rm A}$  when $T_{\rm\parallel\alpha}/T_{\rm\parallel p}=4$, while the threshold for the FM/W instability is about $1.2V_{\rm A}$ \citep{Gary2000a,Li2000}. The temperature anisotropy of the $\alpha$-particles can also significantly reduce the thresholds of parallel A/IC and FM/W instabilities in terms of $U_{\rm p\alpha}$ \citep{Verscharen2013b}.

Only about 2\% of the overall solar-wind fluctuations, in terms of their relative fluctuating power, are compressive  \citep{Chen2016}. Observations show that slow-mode-like fluctuations are a major component among these compressive fluctuations in the solar wind \citep{Howes2012, Klein2012}. Slow-mode-like fluctuations are characterized by an anti-correlation between density and magnetic field strength, a property that these fluctuations share with the magnetohydrodynamics (MHD) slow mode in the low-$\beta_{\rm p}$ limit. A large majority of the compressive fluctuations in the solar wind are found to be quasi-stationary pressure-balanced structures which can be interpreted as highly-oblique slow modes, although the exact orientation of their wavevectors and thus their propagation properties are not fully understood\footnote{We do not invoke non-propagating pressure-balanced structures for our proposed scenario in this article. We only reference to pressure-balanced structures as observational evidence for the slow-mode-like polarization of the compressive fluctuations in the solar wind.} \citep{Bavassano2004, Kellogg2005, Yao2011, Yao2013}. The three-dimensional eddy shape of the compressive fluctuations is highly elongated along the local mean magnetic field \citep{Chen2012}. This wavevector anisotropy ($k_{\perp}/k_{\parallel}$) of the compressive fluctuations is typically greater than six and thus about four times greater than the wavevector anisotropy of the Alfv\'enic fluctuations \citep{Chen2016}, where $k_{\perp}$ and $k_{\parallel}$ are the wavenumbers in the directions perpendicular and parallel to the background magnetic field. Therefore, most compressive fluctuations have an angle of propagation greater than $80^\circ$ with respect to the local mean magnetic field. This result is confirmed by the agreement between MHD predictions of slow-mode polarization properties assuming an angle of $88^\circ$ and observations of the compressive fluctuations in the solar wind
\citep{Verscharen2017}.

%In kinetic theory, two solution branches to the linear Vlasov--Maxwell equations satisfy this polarization property and thus form the kinetic counterparts to the MHD slow mode \citep{Verscharen2017}: ion-acoustic waves and non-propagating pressure balanced structures (PBSs). Multi-scale quasi-static PBSs are pervasive in the solar wind \citep{Tu1994, Bavassano2004, Kellogg2005, Yao2011, Yao2013}, while ion-acoustic waves are more likely damped through Landau damping, especially when $T_{\mathrm e}\lesssim T_{\mathrm p}$ \citep{Barnes1966, Ruan2016}.

Slow-mode-like fluctuations with large amplitudes cause significant changes in the plasma density, temperature, and magnetic field strength. This leads to an effect known as the ``fluctuating-anisotropy effect" \citep{Verscharen2016}. In this framework, slow-mode-like fluctuations quasi-periodically modify the values of $T_{\rm\perp p}/T_{\rm\parallel p}$ and $\beta_{\rm \parallel p}$ at any given point in the plasma, fluctuating around their average values. If these values cross the thresholds for anisotropy-driven kinetic instabilities during this parameter-space trajectory, ion-scale waves grow and scatter the protons, altering the average value of $T_{\rm\perp p}/T_{\rm\parallel p}$. The combined action of compressive fluctuations and the anisotropy-driven instabilities maintains an average value of $T_{\rm\perp p}/T_{\rm\parallel p}$
away from the instability thresholds at a location in parameter space that depends on the amplitude of the large-scale compressive fluctuations.

In this work, we extend the fluctuating-anisotropy effect to a broader ``fluctuating-moment framework'' by incorporating differential flows between protons and $\alpha$-particles. We refer to the modification of $U_{\mathrm p\alpha}$ in the presence of large-scale compressions and the associated amplitude-dependent reduction of the effective thresholds of beam-driven instabilities as the ``fluctuating-beam effect''. We hypothesize that long-wavelength slow-mode-like fluctuations are able to lower the effective thresholds of $\alpha$-particle-driven kinetic instabilities in the solar wind.

In Section~\ref{sec:theory}, we present our  understanding of the framework in which slow-mode-like fluctuations modify the effective instability threshold. We present our results showing the dependence of the effective thresholds of A/IC and FM/W instabilities. In Section~\ref{observe}, we test our theoretical results against solar-wind measurements from the {\it Wind} spacecraft. We present our results and discuss the limitations and further research directions in Section~\ref{sec:sum}.
%($\delta B_{\parallel}/B_0$). We add `0' to the end of the subscript to represent the background or the average value.

\section{Analysis of the fluctuating-beam effect} \label{sec:theory}

\subsection{Conceptual Description}\label{Conceptual Description}

\begin{figure*}
    \plotone{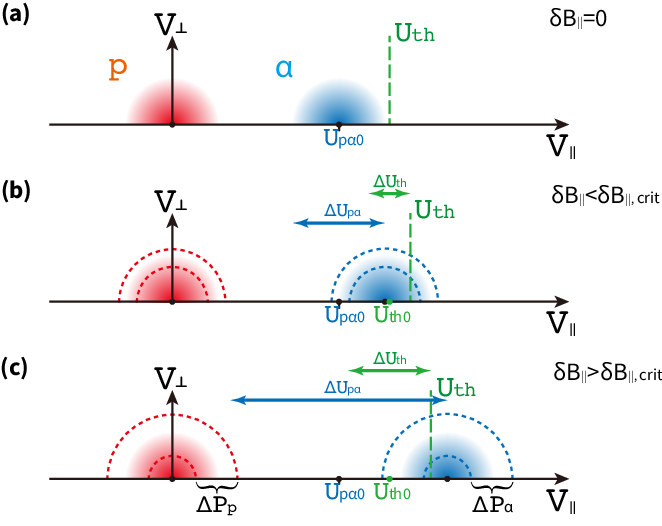}
    \caption{Illustration of the fluctuating-beam effect. We assume the velocity distribution functions of protons (red half-circles) and $\alpha$-particles (blue half-circles) are initially Maxwellian with the same thermal speed. The initial bulk velocity of each component is represented by a black dot. We work in the reference frame of the proton bulk velocity but note that this is not an inertial frame when a slow-mode-like compression is present. (a) Case without slow-mode-like compressions ($\delta B_{\parallel}=0$). Both $U_{\rm p\alpha}$ and $U_{\rm th}$ remain constant, i.e., $U_{\rm p\alpha}=U_{\rm p\alpha0}$ and $U_{\rm th}=U_{\rm th 0}$.  (b) Case with slow-mode-like compressions with an amplitude $0<\delta B_{\parallel}<\delta B_{\parallel,\mathrm{crit}}$. This configuration is still stable throughout all phases of the large-scale compression since $U_{\mathrm p\alpha}<U_{\mathrm{th}}$ at all times. (c) Case with slow-mode-like compressions with an amplitude $\delta B_{\parallel}>\delta B_{\parallel,\mathrm{crit}}$. In this case, $U_{\rm p\alpha}$ crosses the instability threshold $U_{\rm th}$ at least one time instance during the evolution of the large-scale compression. $\Delta U_{\rm p\alpha}$, $\Delta U_{\rm th}$, $\Delta P_{\rm p}$, and $\Delta P_{\rm\alpha}$ represent the variations of the corresponding quantities, where $P_j$ indicates the scalar pressure of species $j$.}
    \label{fig:fig1}
\end{figure*}

Figure~\ref{fig:fig1} illustrates the fundamental concept of the fluctuating-beam effect. In case (a), we assume that no compressive fluctuations are present ($\delta B_{\rm\parallel}=0$) and that $U_{\rm p\alpha 0}$ (i.e., the average $U_{\mathrm p\alpha}$) is below the threshold value $U_{\rm th}$ for any given $\alpha$-particle-driven instability. Since $U_{\rm p\alpha}<U_{\rm th}$ at all times, this case corresponds to an overall stable configuration. 

In case (b), we now include a slow-mode-like fluctuation with a small amplitude $\delta B_{\parallel}\ne 0$. Due to the presence of this fluctuation, all plasma and field parameters participate in the fluctuations according to the polarization properties of the mode. This leads to quasi-periodic fluctuations in $U_{\mathrm p\alpha}$ around $U_{\mathrm p\alpha 0}$ and in the instability threshold $U_{\mathrm{th}}$ around its average value $U_{\mathrm{th} 0}$. These quasi-periodic fluctuations in $U_{\mathrm p\alpha}$ and $U_{\mathrm{th}}$ are in phase (as we see in Section~\ref{Instability thresholds in slow-mode fluctuations}). In case (b), we assume that the amplitude $\delta B_{\parallel}$ is so small that $U_{\rm p \alpha}<U_{\rm th}$ throughout the evolution of the large-scale compression. Therefore, this system remains stable despite the fluctuations in both $U_{\rm p\alpha}$ and $U_{\rm th}$.

In case (c), we now assume that the amplitude $\delta B_{\parallel}$ is so large that $U_{\rm p\alpha}> U_{\rm th}$ for at least a finite time interval during the quasi-periodic evolution of the large-scale compression. We refer to this critical amplitude required to cause $U_{\rm p\alpha}> U_{\rm th}$ at least once as $\delta B_{\parallel,\mathrm{crit}}$. Once $U_{\rm p\alpha}>U_{\rm th}$, A/IC waves or FM/W waves begin to grow due to the excitation of the relevant kinetic instability. We assume the excited instability to grow faster than the timescale of the large-scale compression\footnote{For highly-oblique slow-mode waves at scales comparable to the top end of the inertial range of the solar-wind turbulence, this is a good assumption, even if the maximum growth rate of the instabilities is $\sim 10^{-4}\Omega_{\mathrm p}$, where $\Omega_{\mathrm p}$ is the proton cyclotron frequency.}, so that $U_{\mathrm p\alpha 0}$ is then quickly reduced through wave--particle interactions. We define the amplitude-dependent value of $U_{\mathrm p\alpha 0}$ for which $U_{\mathrm p\alpha}$ reaches $U_{\mathrm{th}}$ once during the large-scale compression as $U_{\rm p\alpha 0, crit}$.
Its value depends on the background plasma conditions and the amplitude of the slow-mode-like compression. In the next section, we quantify the dependence of $U_{\rm p\alpha 0, crit}$ on $\beta_{\rm\parallel p}$ and $\delta B_{\rm \parallel}/B_{\rm 0}$.
% The slow-mode fluctuation adjusts both the differential flow $U_{\rm\alpha}$ and the A/IC instability threshold ($U_{\rm th}$). According to the results of three-fluid theory, the fluctuations of $U_{\rm\alpha}$ and $U_{\rm th}$ are in-phase, which are shown in Figure \ref{fig:fig3}(b) and Figure \ref{fig:fig3}(c). Figure \ref{fig:fig3}(a) shows that if $U_{\rm\alpha0}<U_{\rm th,0}$ in the beginning, the instability would not be excited afterwards when there is no slow-mode fluctuations. When a small-amplitude slow mode wave is present ($\left|\delta B_{\rm\parallel}\right|>0$), $U_{\rm\alpha}$ and $U_{\rm th}$ start the in-phase fluctuation (Figure \ref{fig:fig3}(b)). In this case, the drift velocity fluctuation is not large enough to traverse the instability threshold. As the wave amplitude grows ($\left|\delta B_{\rm\parallel 2}\right|>\left|\delta B_{\rm\parallel 1}\right|$), the amplitude of drift velocity fluctuation grows faster, it can traverse the instability threshold around their maximum values (Figure \ref{fig:fig3}(c)).

\subsection{Slow-mode-like Fluctuations and Their Effect on $\alpha$-particle Beams}

We solve the linear dispersion relation of slow-mode waves with a multi-fluid plasma model to obtain the associated polarization properties for all relevant quantities.  \citet{Xie2014} provides the Plasma Dispersion Relation -- Fluid Version (PDRF) code that solves the linearized multi-fluid equations. We obtain all eigenvalues ($\lambda$) and eigenvectors ($\vec X$) of the linearized multi-fluid eigenvalue problem at a fixed wavenumber ($\vec k$) and with specified background plasma parameters. The number of eigenvalue-eigenvector pairs is $4s+6$ for each calculation, where $s$ is the total number of plasma species. In our case, $s=3$.

For the evaluation of the large-scale compression, we consider a homogeneous plasma consisting of electrons, protons, and $\alpha$-particles with isotropic temperatures ($T_{\parallel j}=T_{\perp j}$) in a uniform background magnetic field $\vec B_0=B_0\hat{\vec e}_z$. We only allow for parallel differential background flows. We ignore viscosity and relativistic effects, so the governing fluid equations are 
\begin{align}
    \frac{\partial n_{ j}}{\partial t} & = -\nabla\cdot\left(n_{ j}{\vec U}_{ j}\right), \\
    \frac{\partial {\vec U}_{ j}}{\partial t} & = -{\vec U}_{ j}\cdot\nabla{\vec U}_{ j}+\frac{q_{ j}}{m_{ j}}\left({\vec E}+\frac{1}{c}{\vec U}_{ j}\times{\vec B}\right)-\frac{\nabla P_j}{n_{ j}m_{ j}}, \\
    \frac{\partial{\vec E}}{\partial t} & = c\nabla\times{\bf B}-4\pi \sum_jq_{ j}n_{ j}{\vec U}_{ j}, \\
    \frac{\partial{\vec B}}{\partial t} & = -c\nabla\times{\bf E},
\end{align}
where $q_{ j}$ is the charge, $m_{ j}$ is the mass, and $P_j$ is the scalar pressure of species $j$; $\vec B$ is the magnetic field; $\vec E$ is the electric field; and $c$ is the speed of light.  We close these equations through the use of the polytropic relation
\begin{equation}
    P_{ j}=C_{ j}n_j^{-\gamma_{j}},
\end{equation}
where $C_{ j}$ is a constant, and $\gamma_{j}$ is the  polytropic index. 
We work in the proton rest frame, which means that $U_{\rm p}=0$. Given the background values $n_{\rm p0}$, $n_{\rm\alpha0}$, and ${\vec U}_{\rm p\alpha0}$, we obtain $n_{\rm e0}$ and ${\vec U}_{\rm e0}$  from the current-free and charge-neutrality conditions.

We assume highly oblique propagation for our slow mode and set $\theta\left({\vec k},{\vec B}_0\right)=88^\circ$, where $\theta\left({\vec k},{\vec B}_0\right)$ is the angle between the wavevector ${\vec k}$ and ${\vec B}_0$. We focus on the parameter range  $0<U_{\rm p\alpha0}/V_{\rm A}<1.5$ and $0.1<\beta_{\rm \parallel p0}<10$. This range covers the majority of solar-wind conditions at 1\,au \citep{Maruca2011}. We list all background parameters of our calculation in Table~\ref{tab:tab1}. 

In the inner heliosphere, $\gamma_\mathrm{e}\sim 1.18$ based on Parker Solar Probe data \citep{Abraham2022}. Observations suggest that $\gamma_{\mathrm{p}}\sim 5/3$ as expected for an adiabatic, mono-atomic gas \citep{Huang2020,Nicolaou2020}, and $\gamma_{\alpha}\sim1.61$ \citep{Durovcova2019}. The observed polytropic indexes for $\alpha$-particles are generally different for the parallel ($\gamma_{\parallel\alpha}$) and perpendicular ($\gamma_{\perp \alpha}$) pressures, and depend on the solar-wind speed \citep{Durovcova2019}. In fast wind, $\gamma_{\parallel \alpha}\sim1.65$ and $\gamma_{\perp \alpha}\sim1.57$. In slow wind, $\gamma_{\parallel \alpha}\sim1.05$ and $\gamma_{\perp \alpha}\sim1.24$. Since the largest $U_{\mathrm p\alpha}$ are observed in fast wind -- making fast wind more relevant for our study --, we choose the average fast-wind polytropic index  $\gamma_{\alpha}\sim1.61$ in our analysis below. We discuss the dependence of our results on the particular choice of polytropic indexes in Appendix~\ref{appendix}. The density ratios and temperature ratios are set according to representative values measured in the solar wind \citep[see][]{Kasper2008, Maruca2013}. 

Our PDRF analysis confirms that, in slow modes, all plasma and field quantities fluctuate, including $n_{ j}$, $T_{ j}$, $\vec U_{j}$, and $\vec B$. This allows us to quantify the variations of $U_{\mathrm p\alpha}$ and $U_{\mathrm{th}}$ to determine $U_{\mathrm p\alpha,\mathrm{crit}}$. 

\begin{deluxetable*}{ccccccc}
\tablenum{1}
\tablecaption{Assumed plasma parameters for our evaluation of the linearized multi-fluid equations. \label{tab:tab1}}
\tablewidth{0pt}
\tablehead{
\colhead{$j$} & \colhead{$q_{ j}$} & \colhead{$m_{ j}/m_{\rm p}$} & \colhead{$n_{ j0}/n_{\rm p0}$} & \colhead{$T_{\rm\perp j0}/T_{\rm\parallel j0}$} & \colhead{$T_{\rm\parallel j0}/T_{\rm\parallel p0}$} & \colhead{$\gamma_{ j}$} 
}
\startdata
$\mathrm p$ & $+1$ & 1 & 1 & 1 & 1 & 5/3 \\
$\alpha$ & $+2$ & 4 & 0.05 & 1 & 4 & 1.61  \\
$\mathrm e$ & $-1$ & 1/1836 & 1.1 & 1 & 1 & 1.18 
\enddata
\end{deluxetable*}

%\section{The Dispersion Relations of Quasi-Perpendicular-Propagating Slow Modes} \label{sec:disp}

\subsection{Thresholds for A/IC and FM/W Instabilities Depending on Slow-mode Amplitude} \label{Instability thresholds in slow-mode fluctuations}

\citet{Verscharen2013b} derive two analytical expressions for the parallel A/IC and FM/W instabilities driven by the relative drift between $\alpha$-particles and protons. According to their derivation, the A/IC instability is excited when
\begin{equation}\label{eq1}
    U_{\rm p \alpha}>\frac{1}{2}\left[\sigma_1w_{\rm p}+V_{\rm A}+\sqrt{\left(V_{\rm A}+\sigma_1w_{\rm p}\right)^2-2V_{\rm A}^2}\right]-\sigma_1w_{\rm p},
\end{equation}
where $\sigma_1=2.4$ and $w_{j}\equiv\sqrt{2k_{\mathrm B}T_{j}/m_j}$ is the thermal speed of species $j$. The condition for driving of the FM/W instability is
\begin{equation}\label{eq2}
    U_{\rm p\alpha}>V_{\rm A}+\frac{V_{\rm A}^2}{4\sigma_2w_{\rm\alpha}},
\end{equation}
where $\sigma_2=2.1$. The dimensionless parameters $\sigma_1$ and $\sigma_2$ are empirical and based on comparisons between the analytical expressions and solutions of the linear Vlasov--Maxwell dispersion relation with $\gamma_{\rm m}=10^{-4}\Omega_{\rm p}$ when $n_{\rm\alpha}/n_{\rm p}=0.05$ and $T_{\rm\alpha}=4T_{\rm p}=4T_{\rm e}$, where $\Omega_{\rm p}$ is the proton cyclotron frequency and $\gamma_{\rm m}$ is the maximum growth rate of the instabilities.
Since we consider isotropic temperatures only, the slow-mode fluctuations impact the instability thresholds by modifying $w_{\rm p}$, $w_{\alpha}$, $\left|\vec B\right|$ and $n_{\rm p}$.

Our three-fluid model predicts that the relative amplitudes and the phase differences of $ w_{\rm p}$ and $ n_{\rm p}$ to $\delta B_{\parallel}$ do not depend on  $k_{\parallel}d_{\rm p}$ when  $k_{\parallel}d_{\rm p}\lesssim 10^{-2}$ (not shown here), where $d_{\mathrm p}\equiv V_{\mathrm A}/\Omega_{\mathrm p}$ is the proton inertial length. Hence, we extend the applicability of our results to  all slow-mode waves at $k_{\parallel}d_{\rm p}\ll 10^{-2}$. We choose solutions at $k_{\parallel}d_{\rm p}=3\times10^{-4}$ to conduct our further analysis. 

Since our assumed background density and temperature ratios are the same as those used by \citet{Verscharen2013b}, we directly use Equations~(\ref{eq1}) and~(\ref{eq2}) to study the variation of the instability thresholds driven by slow-mode waves. While evaluating these expressions, we ensure that the fluctuating temperatures $T_{j}$ and  densities $n_{ j}$ of all species remain positive during a full slow-mode wave period.

\subsection{Determination of the Effective Instability Thresholds}\label{results}

From the dispersion relation, we construct the time series of each quantity $Q\left(t\right)=Q_{\rm 0}+\delta Q(t)$, assuming a plane-wave fluctuation of all quantities with $\delta Q(t)=\mathrm{Re}(\tilde{Q}e^{-i\omega t})$, where $\tilde{Q}$ represents the complex amplitude of quantity $Q$, $\omega$ is the wave frequency, and $\mathrm{Re}(\cdot)$ extracts the real part of a complex variable. We calculate the time series of $U_{\rm p\alpha}(t)$ and $U_{\rm th}(t)$. In general, $\delta Q(t)$ is in phase or anti-phase with $\delta B_{\parallel}(t)$. 
By analyzing $U_{\rm p\alpha}\left(t\right)$ and $U_{\rm th}\left(t\right)$ as functions of $t$ and $\delta B_{\rm\parallel}/B_{\rm 0}$, we determine $\delta B_{\rm\parallel,crit}/B_{\rm 0}$ depending on $U_{\rm p \alpha0}/V_{\rm A 0}$ and $\beta_{\rm \parallel p0}$, where $V_{\rm A 0}$ and $\beta_{\rm \parallel p0}$ represent the values of $V_{\rm A}$ and $\beta_{\rm\parallel p}$ in the limit $\delta B_{\parallel}\rightarrow 0$.

%For particular $U_{\rm\alpha0}/V_{\rm A}$ and $\beta_{\rm p\parallel}$, $U_{\rm\alpha}$ exceeds $U_{\rm th}$ at least for a finite time (phase) interval during slow-mode-like perturbation when $\delta B_{\rm\parallel}>\delta B_{\rm\parallel,crit}$. On the contrary, $U_{\rm\alpha}(t)$ is always smaller than $U_{\rm th}(t)$ when $\delta B_{\rm\parallel}<\delta B_{\rm\parallel,crit}$, although they both fluctuate according to the slow-mode polarization. Since we hypothesize that, 

%Once $U_{\rm p\alpha}$ is greater than $U_{\rm th}$, $U_{\rm\alpha}$ is then rapidly reduced by kinetic instability in a fast time scale compared to the slow-mode wave period, the average value of $U_{\rm\alpha}$ is thereby diminished. Hence, we can define a critical $U_{\rm\alpha 0}$, $U_{\rm\alpha 0,crit}$ which we also refer to as effective threshold. When $U_{\rm\alpha 0} > U_{\rm\alpha 0,crit}$, the plasma system is unstable to the instability at least for a finite time interval and $U_{\rm\alpha0}$ is consequently reduced to the stable regime through wave-particle interaction. On the contrary, if $U_{\rm\alpha 0}<U_{\rm\alpha 0, crit}$, although both $U_{\rm\alpha}$ and $U_{\rm th}$ are fluctuating according to the slow-mode polarization, $U_{\rm\alpha}$ is always smaller than $U_{\rm th}$ during the evolution.

\subsubsection{Fluctuating-beam effect for the A/IC Instability} 

Figure~\ref{fig:fig2} shows the dependence of $U_{\rm p\alpha 0,crit}$ on $\beta_{\rm\parallel p0}$ and $\delta B_{\rm\parallel}/B_0$ for the A/IC instability. When there is no slow-mode wave ($\delta B_{\rm\parallel}/B_0=0$), the A/IC instability threshold changes slowly from $\sim0.7V_{\rm A0}$ at $\beta_{\rm\parallel p0}=0.1$ to $\sim0.95V_{\rm A0}$ at $\beta_{\rm\parallel p0}=10$. This case corresponds to the classical limit in a homogeneous plasma without large-scale compressions. 

When a small-amplitude slow-mode wave ($\delta B_{\rm\parallel}/B_0=0.01$) is present, $U_{\rm p\alpha 0,crit}$ decreases compared to the classical case. The decrease of  $U_{\rm p\alpha 0,crit}$ is more significant at $\beta_{\rm \parallel p0}\lesssim 1$. However, $U_{\rm p\alpha 0,crit}$ is still greater than 0.85$V_{\rm A 0}$ at $\beta_{\rm\parallel p0}>5$ even at large wave amplitudes  ($\delta B_{\rm\parallel}/B_{\mathrm{0}}\sim 0.2$). Our solutions break down at low $\beta_{\rm\parallel p0}$ first, as the wave amplitude increases. These are parameter regimes, in which our linear framework produces negative densities or temperatures, which is unphysical. We mark these break-down points as black dots at the low-$\beta_{\mathrm\parallel p0}$ end of our solution plots.

\subsubsection{Fluctuating-beam effect for the FM/W Instability} 

\begin{figure}
   
    \plotone{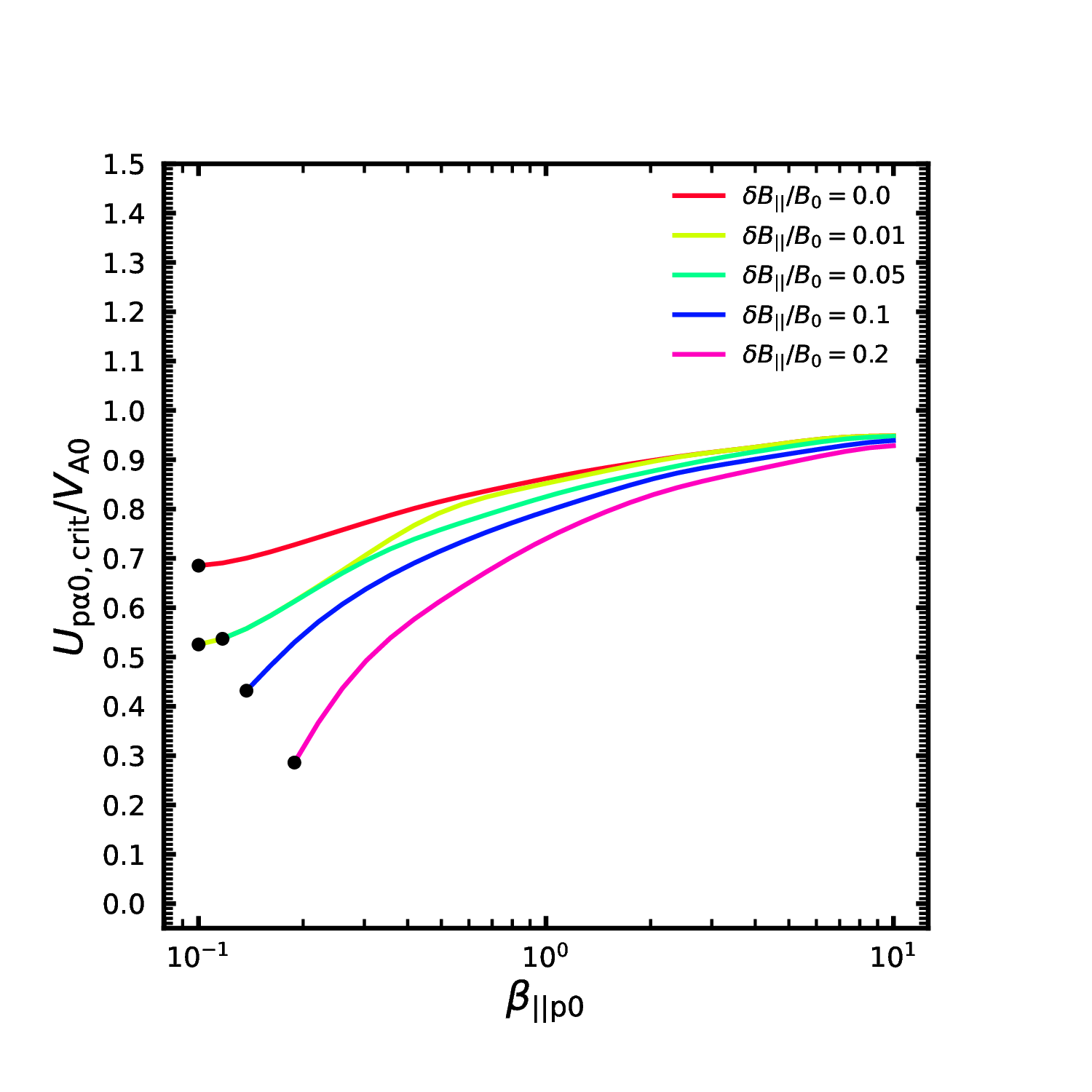}
    \caption{Dependence of the maximum permitted $U_{\rm p\alpha 0}$ in terms of $U_{\rm p\alpha 0,crit}/V_{\rm A0}$ on $\beta_{\rm \parallel p0}$ and $\delta B_{\rm\parallel}/B_0$ for the A/IC instability. The normalized amplitude $\delta B_{\rm\parallel}/B_0$ varies between  0 to 0.2.}
    \label{fig:fig2}
\end{figure}

\begin{figure}
 
\plotone{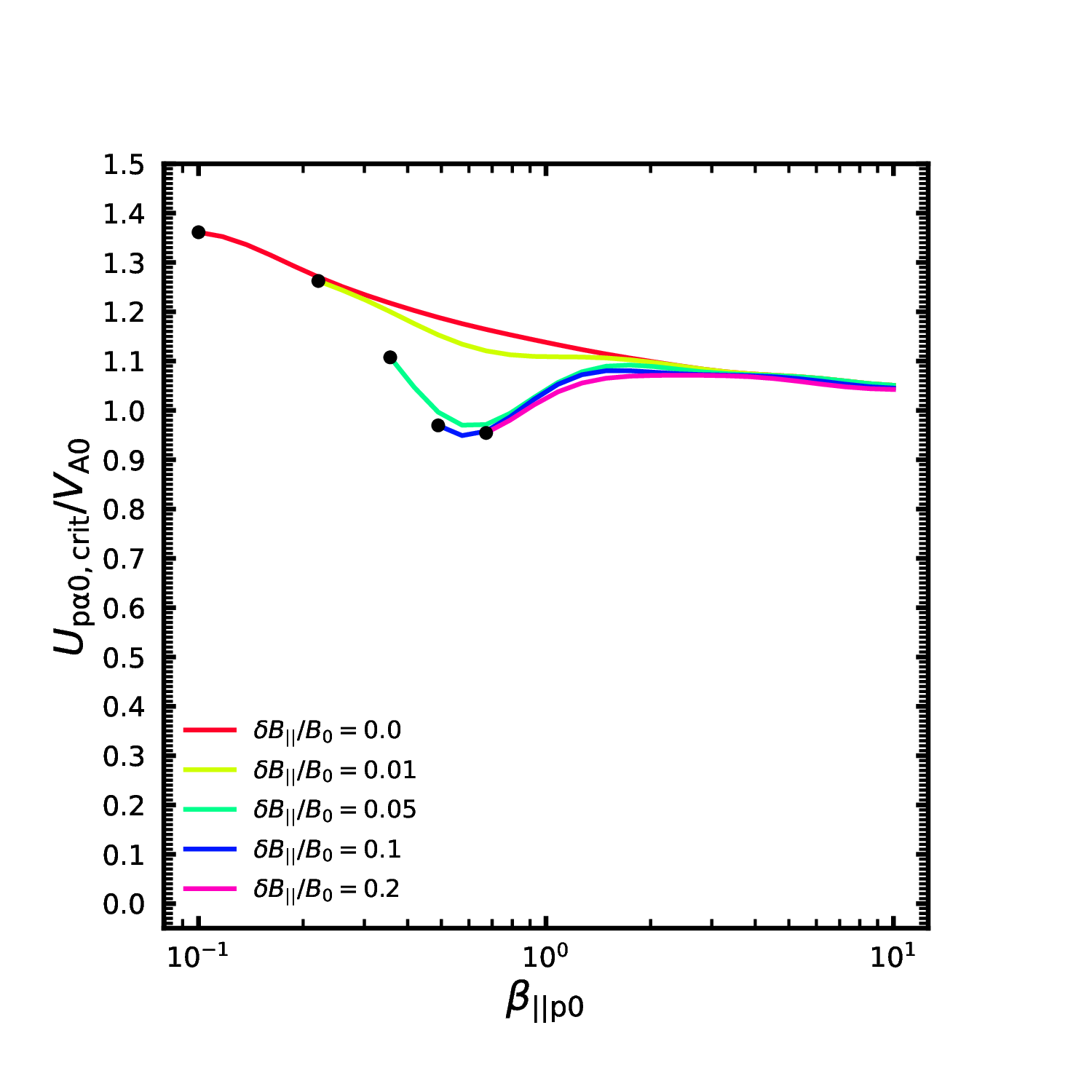}
    \caption{Dependence of the maximum permitted $U_{\rm p\alpha 0}$ in terms of $U_{\rm p\alpha 0,crit}/V_{\rm A0}$ on $\beta_{\rm\parallel p0}$ and $\delta B_{\rm\parallel}/B_0$ for the FM/W instability. The normalized amplitude $\delta B_{\rm\parallel}/B_0$ varies between 0 to 0.2.}
    \label{fig:fig3}
\end{figure}

Figure~\ref{fig:fig3} has the same format as Figure~\ref{fig:fig2} but shows the FM/W instability thresholds. Unlike the case of the A/IC instability, the effective FM/W instability threshold decreases from $\sim1.35V_{\rm A0}$ at $\beta_{\rm\parallel p0}\sim0.1$ to $\sim1.05V_{\rm A0}$ at $\beta_{\rm \parallel p0}\sim10$ when $\delta B_{\parallel}/B_{ 0} =0$. The slow-mode fluctuations mainly affect the FM/W instability thresholds at $\beta_{\rm\parallel p0}<2$. $U_{\rm p\alpha 0, crit}$ is reduced as the wave amplitude increases. At $\beta_{\rm\parallel p0}>2$, the curves in our figure approximately overlap as long as our method provides physical solutions. 

%$U_{\rm p\alpha 0, crit}$ sharply decreases from 0.8 at $\beta_{\rm\parallel p0}\sim5$ to 0 at $\beta_{\rm\parallel p0}\sim1.5$, which indicates that the fluctuating-beam effect does not permit any relative drift between protons and $\alpha$-particles in this parameter range. We note, however, that waves with $\delta B_{\rm\parallel}/B_{\rm 0}=1.0$ clearly lie beyond the linear regime for which our PDRF solutions apply. 

%The amplitude ratio $\left|\delta U_{\rm p\parallel}\right|/\left|\delta U_{\rm\alpha\parallel}\right|$ is greater than 1 when $U_{\rm\alpha 0}=0$. This ratio decreases with increasing $U_{\rm\alpha 0}$ and transits to values less than 1 at about $U_{\rm\alpha 0}=0.01V_{\rm A}$ in the $\beta_{\rm p\parallel}$ range analyzed here. Therefore, $U_{\rm\alpha}\left(t\right)$ fluctuates out-of-phase with $U_{\rm th}\left(t\right)$ for $U_{\rm\alpha 0}=0$ and in-phase with $U_{\rm th}\left(t\right)$ for $U_{\rm\alpha 0}\ne 0$. This phase relation lowers the effective threshold for $\delta B_{\rm\parallel}/B_{\rm 0}=1$.

\section{Comparison with Observations}\label{observe}

We compare our theoretical predictions from Section~\ref{sec:theory} with measurements of protons, $\alpha$-particles, and compressive fluctuations in the solar wind recorded by the {\it Wind} spacecraft at 1~au. We use data from the two Faraday cups which belong to the Solar Wind Experiment \citep[SWE;][]{Ogilvie1995}. They generate an ion spectrum with $\sim$90~s cadence in most situations. We base our analysis on the non-linear fitting routine provided by \citet{Maruca2012}. This procedure provides us with the plasma density, velocity, and temperature for both protons and $\alpha$-particles. It incorporates magnetic-field measurements with a cadence of 3~s from the Magnetic Field Investigation experiment \citep[MFI;][]{Lepping1995} to determine the parallel and perpendicular components of temperature and velocity. The procedure also provides the differential flow $U_{\mathrm p\alpha}$ between $\alpha$-particles and protons, assuming it aligns with the magnetic-field direction. This dataset spans from 1994 November 21 to 2010 July 26 and
contains 2\,148\,228 ion spectra, after removing measurements near the Earth's bow shock or with poor data quality. We divide the dataset into non-overlapping 1-hour intervals. We exclude all intervals with data gaps greater than 20~min from further analysis. 

For each 1-hour interval, we calculate the average values $n_{\rm p0}$, $T_{\rm p 0}$, $U_{\rm p\alpha 0}$, $V_{\rm A 0}$, and $B_0$. We then obtain the average $\beta_{\rm\parallel p0}$ and the average $U_{\rm p\alpha 0}/V_{\rm A0}$ from these averages. For the amplitude of the compressive fluctuations, we use the root-mean-squared (rms) value of $\delta\left|{\vec B}\right|$, denoted as $\delta\left|{\vec B}\right|_{\rm rms}$, instead of $\delta B_{\rm\parallel}$, to avoid the impact of inaccuracies in the determination of the parallel magnetic field direction. We bin the data into eight $\beta_{\rm\parallel p0}$ ranges from 0.1 to 10 on a logarithmic scale. For each $\beta_{\rm\parallel p0}$ bin, we divide the sub-dataset into 30 bins of $\delta\left|{\vec B}\right|_{\rm rms}/B_0$ in the range from 0.0 to 1.0. We then calculate the probability density function (PDF) of $U_{\rm p\alpha 0}/V_{\rm A0}$ for each $\delta\left|{\vec B}\right|_{\rm rms}/B_0$ bin. 

\begin{figure*}
\plotone{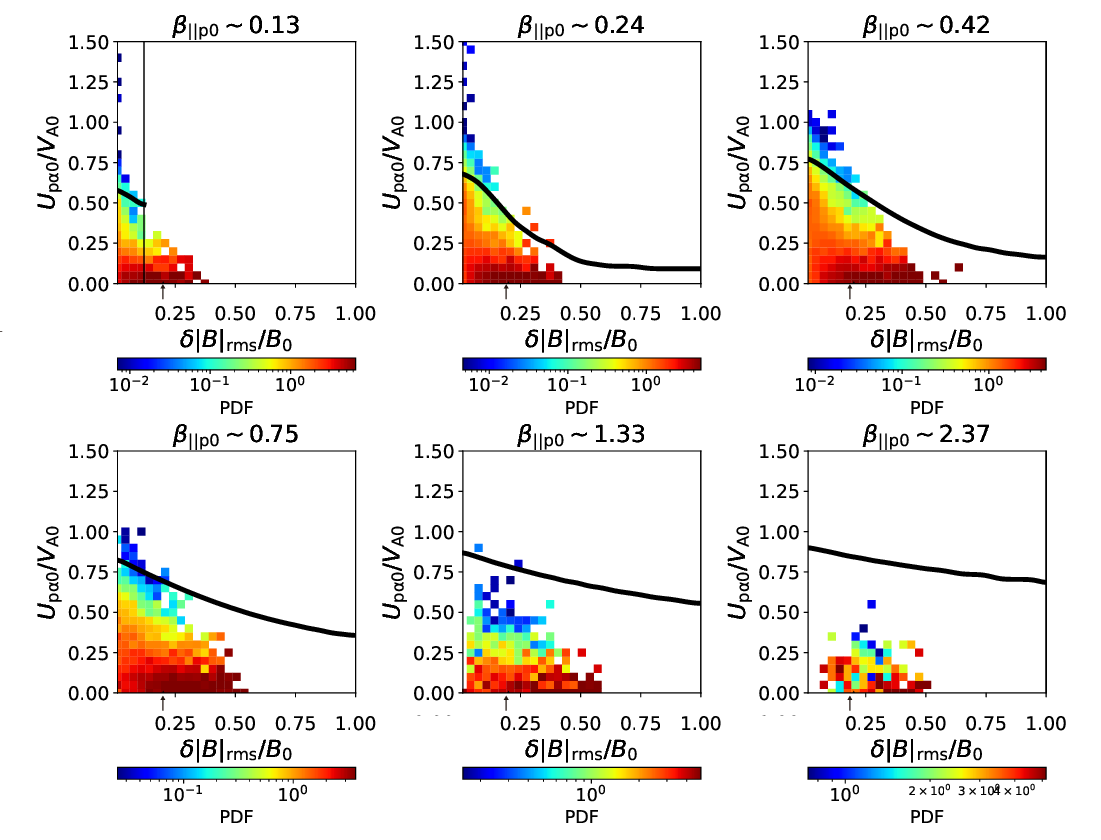}
    \caption{Comparison of our theoretical predictions for the effective instability thresholds $U_{\mathrm p\alpha 0,\mathrm{crit}}$ with observations from \textit{Wind}. We show the comparison for six different $\beta_{\rm\parallel p0}$ values. The color-coding represents the binned PDFs of 1-hour averaged $U_{\mathrm p\alpha 0}/V_{\rm A0}$ for each $\delta |\vec B|_{\mathrm{rms}}/B_0$ bin. The thick solid black curves show the theoretical effective thresholds $U_{\rm p\alpha 0,crit}$ for the A/IC instability from Section~\ref{sec:theory}. The vertical black line marks the amplitude above which our linear model breaks down, which is only the case in the first panel. The arrows in each panel below the abscissa indicate the amplitude $\delta |\vec B|_{\mathrm{rms}}/B_0=0.2$, the approximate limit of validity of the linear model in all panels.}
    \label{fig:fig4}
\end{figure*}

Figure~\ref{fig:fig4} shows the PDF of solar-wind data in the $U_{\mathrm p\alpha 0}/v_{\mathrm A0}$-$\delta|\vec B|_{\mathrm{rms}}/B_0$ plane for six different $\beta_{\parallel\mathrm p0}$-ranges. We also overplot the effective instability threshold $U_{\rm p\alpha 0, crit}/v_{\mathrm A0}$ (black curve) for the A/IC instability. 
The parameter space below the effective instability threshold represents the stable region. The PDF of the measured data is largely confined to this stable parameter space in all cases shown. Only an insignificant number of points are beyond the thresholds. We find that $U_{\mathrm p\alpha 0,\mathrm{crit}}/V_{\rm A0}$  decreases with $\delta |\vec{B}|_{\mathrm{rms}}/B_0$, and the PDFs of our measured data broadly follow this trend. $U_{\rm p\alpha 0, crit}$ bounds the data well in the stable region, especially for $\beta_{\rm\parallel p0}<0.5$. At $\beta_{\rm\parallel p0}\sim 0.75$, the decrease of the maximum $U_{\rm p\alpha 0}/V_{\rm A0}$ is steeper than the decrease in $U_{\rm p\alpha 0, crit}$ while in the high $\beta_{\rm \parallel p0}$-range, the data points locate farther away from the $U_{\rm p\alpha 0, crit}$-curve. 
This finding  is consistent with our concept that large-scale slow-mode-like fluctuations regulate $U_{\rm p\alpha 0}$ through the fluctuating-beam effect. Some data points lie in the parameter space in which our linear model breaks down when $\beta_{\rm\parallel p0}\sim0.13$, which is indicated by the vertical line. We emphasize that our results are strictly only valid for $\delta |\vec {B}|_{\mathrm{rms}}/B_0\ll 1$, when the linear assumption is valid. We mark a relative amplitude of $\delta |\vec{B}|_{\mathrm{rms}}/B_0= 0.2$ with a black arrow in each panel in Figure~\ref{fig:fig4} and Figure~\ref{fig:fig7} (see Appendix \ref{appendix}) to indicate a reasonable limit for the application of our model.

\section{Discussion and Conclusions} \label{sec:sum}

We introduce the new concept of the ``fluctuating-beam effect" as part of a broader ``fluctuating-moment framework''. The fluctuating-beam effect describes a link between large-scale compressive fluctuations and small-scale kinetic instabilities driven by relative drifts between ion components in collisionless plasmas. In the solar wind, long-period slow-mode-like fluctuations  modulate the background plasma parameters, including $n_{j}$, $T_{j}$, $B$, $U_{\rm p \alpha}$, and $U_{\rm th}$. Once $U_{\rm p \alpha 0}$ is greater than the amplitude-dependent critical value $U_{\mathrm p\alpha0,\mathrm{crit}}$, this slow-mode modulation creates unstable conditions for the growth of beam-driven kinetic microinstabilities. These microinstabilities generate waves on a timescale much shorter than the period of the slow-mode-like fluctuations. The separation of timescales is very large in the solar wind as we show in the following. We estimate the typical slow-mode wave frequency of waves with a wavelength comparable to the top end of the inertial range of solar-wind turbulence at heliocentric distances of about 1~au  as \citep{Tu1995, Bruno2013}
\begin{equation}
\omega_\mathrm{s}\sim k_\mathrm{\parallel}d_\mathrm{p}\sqrt{\beta_\mathrm{p}}\Omega_\mathrm{p}\sim 10^{-6}\Omega_\mathrm{p},
\end{equation}
where $k_\mathrm{\parallel}d_\mathrm{p}\sim10^{-6}$ and 0.1$<\beta_{p}<$10 under typical plasma conditions for our study. The typical relevant maximum growth rate of the beam-driven instabilities is 
\begin{equation}\gamma_{\mathrm m}\sim10^{-4}-10^{-3}\Omega_\mathrm{p}
\end{equation}
according to observations \citep{Klein2018} and simulations \citep{Klein2017}. Therefore, we find
\begin{equation}\label{scalesep}
\gamma_\mathrm{m}\gg \omega_\mathrm{s}.
\end{equation} 
Equation~(\ref{scalesep}) is a critical condition for the applicability of our model.

The scattering of $\alpha$-particles by the unstable waves is the main contributor to the reduction of the local $U_{\rm p \alpha}$ and hence to the reduction of the average $U_{\rm p\alpha 0}$. This process maintains $U_{\rm p \alpha 0}$ well below  the classical (average) instability thresholds $U_{\mathrm{th}0}$ in a homogeneous plasma. The difference between $U_{\mathrm{th}0}$ and $U_{\mathrm p\alpha 0,\mathrm{crit}}$ depends on the amplitude of the compressive fluctuations. The effective threshold $U_{\mathrm p\alpha 0,\mathrm{crit}}$ decreases with increasing $\delta B_{\parallel}/B_0$. If $U_{\mathrm p\alpha 0}>U_{\rm p\alpha 0,crit}$, the plasma becomes unstable at least for a portion of each compressive fluctuation period. Otherwise, if $U_{\mathrm p\alpha 0}<U_{\rm p\alpha0,crit}$, the plasma is stable throughout the evolution of the slow-mode compression, even though the instantaneous $U_{\mathrm p\alpha}$ varies.

We quantify the dependence of $U_{\rm p\alpha 0,crit}$ on the background plasma parameters and the compression amplitude for A/IC and FM/W instabilities.
%We consider a plasma consisting of electrons, protons, and $\alpha$-particles with isotropic temperatures and calculate the dispersion relations and transport ratios of long-wavelength quasi-perpendicular slow-mode waves with $\theta\left(\bf{k},\bf{B_0}\right)=88^\circ$. We quantitatively study the distribution of $U_{\rm\alpha 0, crit}$ as a function of $\delta B_{\rm\parallel}/B_{\rm 0}$ and $\beta_{\rm p\parallel}$. 
For the A/IC instability, $U_{\rm p\alpha 0,crit}$ decreases with increasing slow-mode wave amplitude. For the FM/W instability, $U_{\rm p\alpha 0,crit}$ is  significantly reduced when $\beta_{\rm\parallel p0}\lesssim 1$ compared to $U_{\mathrm{th}0}$. We note, however, that the validity of our linear model for the slow-mode-like  fluctuations is questionable even at smaller amplitudes.

Although solar-wind turbulence is a nonlinear process, its fluctuations display many properties that are similar to the properties of linear waves \citep{He2011, Chen2013, Howes2012}. A large body of observational work \citep{Yao2011, Klein2012, Howes2012, Chen2012, Verscharen2017, Safrankova2021} suggests the presence of slow-mode-like fluctuations in the solar wind, although the origin of these fluctuations is still unclear. According to linear theory, classical slow-mode waves are strongly damped in a homogeneous Maxwellian plasma, especially at intermediate to high $\beta_\mathrm{p}$ and when $T_{\mathrm e}\lesssim T_{\mathrm p}$ \citep{Barnes1966}. Their confirmed existence in the solar wind is thus a major outstanding puzzle in space physics. Possible explanations for this discrepancy between theory and observations include the passive advection of slow-mode-like fluctuations through the dominant Alfv\'enic turbulence \citep{Schekochihin2009, Schekochihin2016}, the fluidization of compressive plasma fluctuations by anti-phase-mixing \citep{Parker2016, Meyrand2019}, or the suppression of linear damping through effective plasma collisionality by particle scattering on microinstabilities \citep{Riquelme2015, Verscharen2016}.

Our comparisons of the effective instability thresholds of the A/IC instability with  measurements from the {\it Wind} spacecraft show that the effective instability thresholds bound the distribution of observed $U_{\rm p\alpha 0}$-values well. This result is consistent with our predictions for the fluctuating-beam effect.  We use this observation to justify our use of linear multi-fluid theory to the large-scale compressions in the solar wind. 

Our results are largely independent of the averaging time for the definition of the background drift speed $U_{\mathrm p\alpha 0}$. To illustrate this point, we make the following comparison. In our dataset, the average $d_{\mathrm p}$ is about 98~km, and the average solar-wind speed is about 426\,km/s. Therefore, the convection time of slow-mode waves with  $k_\mathrm{\parallel}d_{\mathrm p}\sim3\times10^{-4}$ is about 80\,min. Under typical solar-wind conditions ($0.1<\beta_\mathrm{\parallel \mathrm p0}<10$ and $0<U_\mathrm{p\alpha 0}/V_\mathrm{A}<1.5$), the phase speed for these slow-mode waves ranges from $\sim 0.4V_{\mathrm{A}}$ to $\sim V_{\mathrm{A}}$ based on multi-fluid calculations. Hence, the periods of these waves are between $\sim 11.6\,\mathrm h$ and $\sim 29\,\mathrm h$. The characteristic growth time for the instability is about 16\,min. Therefore, during the one-hour averaging interval, the spacecraft approximately measures 1/20-th wave period on average. During this phase, the instability has sufficient time to reduce the local drift velocity of the $\alpha$-particles, as required by our set of assumptions.
%As the slow-mode wave amplitude increases, the required $U_{\rm\alpha 0}/V_{\rm A}$ to drive A/IC instability is reduced, which is more significant at low $\beta_{\rm p}$ range ($\beta_{\rm p\parallel}<1$). However, the thresholds for FM/W instability are nearly impervious except for $\beta_{\rm p\parallel}<5$ when $\delta B_{\parallel}/B_0=1$. The $\beta_{\rm p\parallel}$ range for which the linear assumption is violated increases with wave amplitude and extends from lower $\beta_{\rm p\parallel}$ to higher $\beta_{\rm p\parallel}$ when $\delta B_{\parallel}/B_0\le0.5$. It is worth noting that both A/IC and FM/W instability thresholds decrease to zero at $\sim\beta_{\rm p\parallel}=1.05$ before the linear assumption is violated if we trace from larger $\beta_{\rm p\parallel}$ to lower $\beta_{\rm p\parallel}$ for the lines of $\delta B_{\rm\parallel}/B_0=1$.

Our model for the slow-mode-like compressions is based on the linearized multi-fluid equations. This linearization is only valid when $\delta B_{\parallel}/B_0\ll 1$. Nevertheless, linear predictions often describe the solar wind well, even when the fluctuation amplitude is greater than justifiable based on the linear assumption \citep{Howes2012, Chen2012, Verscharen2017}.  To some degree, the good agreement between our model predictions and our observations in Figure~\ref{fig:fig4} supports our application of linear theory beyond the very-small-amplitude limit \emph{a posteriori}. We emphasize at this point, however, that our framework based on linear predictions is strictly applicable only when $\delta B_{\parallel}/B_0\ll 1$.

The slow-mode-like compressions provide a self-consistent parallel electric field $E_{\parallel}\sim\nabla P_{\mathrm e}$. In our instability analysis, we implicitly treat this large-scale electric field as constant due to the large spatial and temporal scale separation between the slow-mode wave and the instability. Such a constant $E_{\parallel}$ impacts the (assumed to be constant) bulk speed of the local background plasma for the instabilities. In our instability analysis, the local background bulk velocity of the plasma is defined as the superposition of the global background plasma velocity vector and the local bulk speed fluctuation vector of the slow-mode wave. We evaluate the instability criteria in the reference frame in which the local background bulk velocity is zero. This choice of reference frame does not affect the physics of the kinetic instability.

The instability thresholds and thus the fluctuating-beam effect depend on more parameters than we can explore in this work. For example, the thresholds for the A/IC and FM/W instabilities depend on $T_{\rm\perp\alpha}/T_{\rm\parallel\alpha}$, $\beta_{\rm \parallel p}$, and $\beta_{\rm\parallel\alpha}$ \citep{Verscharen2013b}. The thresholds for both instabilities have a modest dependence on $n_{\rm\alpha}/n_{\rm p}$ in the range of 0.01 to 0.05, which is typical in both the slow wind and fast wind \citep{Bame1977,Kasper2007}.  Two-dimensional simulations of the FM/W instability under solar-wind conditions show that, when $U_{\rm p\alpha0}/V_{\rm A0}\approx 1.5$, $\gamma_{\rm m}$ does not change with $T_{\rm\parallel e}/T_{\rm\parallel p}$ and only slightly decreases as $T_{\rm\perp p}/T_{\rm\parallel p}$ and $T_{\rm\perp\alpha}/T_{\rm\parallel\alpha}$ increase from 0.5 to 1 \citep{Gary2000a}. The dependence of our results on these other plasma parameters merits future investigation.

Beams and temperature anisotropies often occur together. For example, hybrid simulations of $\alpha$-particle-driven drift instabilities \citep{Ofman2007,Ofman2022} show that a Maxwellian $\alpha$-particle population with sufficiently large drift speed drives A/IC instabilities, which then create temperature anisotropy in the $\alpha$-particle population through velocity-space diffusion. During the nonlinear evolution of this combined kinetic process, the $\alpha$-particles develop a marginally stable temperature anisotropy, and the drift speed decreases.
Plasma turbulence is also capable of generating temperature anisotropy in the $\alpha$-particle component. However, simulations suggest that the instability-induced generation of temperature anisotropy is more efficient than the turbulence-induced generation of temperature anisotropy in the $\alpha$-particle component \citep{Ofman2007}.
This complex interplay between various kinetic processes highlights that a general fluctuating-moment framework is necessary that includes the impact of large-scale fluctuations on all kinetic microinstabilities.

%We emphasize that considering different background plasma conditions will be substantial in further research to advance our understanding of the fluctuating-beam effect, such as temperature anisotropy, a different density or temperature ratio between different species and different adiabatic coefficients. 

The fluctuating-beam effect is a consequence of the turbulent nature of the solar wind. This inhomogeneity leads to lower effective instability thresholds compared to homogeneous plasmas. We anticipate that a similar fluctuating-beam effect also applies to proton-beam-driven instabilities. Through the lower effective instability thresholds, kinetic microinstabilities are likely to remove more free energy from the $\alpha$-particle beams globally than previously expected, so that instabilities are probably even more important than previously assumed \citep{Verscharen2015}.

\begin{acknowledgments}
D.V.~is supported by STFC Ernest Rutherford Fellowship ST/P003826/1. D.V.~and C.J.O.~are supported by STFC Consolidated Grants ST/S000240/1 and ST/W001004/1. The work at Peking University is supported by NSFC under contract Nos. 42174194 and 41874200, and by CNSA under contract Nos. D020301, D020302, and D050601, as well as National Key R\&D Program of China No. 2021YFA0718600. 
- This research was supported by the International Space Science Institute (ISSI) in Bern, through ISSI International Team project \#463 (Exploring The Solar Wind In Regions Closer Than Ever Observed Before) led by L.~Harrah and through ISSI International Team project \#563 (Ion Kinetic Instabilities in the Solar Wind in Light of Parker Solar Probe and Solar Orbiter Observations) led by L. Ofman and L.~Jian.
\end{acknowledgments}

\appendix \label{appendix}
\section{Dependence on Polytropic indexes}

Plasma descriptions typically fall in one of two categories: fluid or kinetic descriptions. These two approaches do not necessarily lead to the same results even at large fluid scales due to the assumed pressure closure of the fluid equations. The compressive behavior of slow-mode-like fluctuations at large scales depends on the choice of the effective polytropic indexes $\gamma_j$ of the involved plasma species. For large-scale kinetic slow modes, the effective polytropic indexes are $\gamma_\mathrm{e}=1$, $\gamma_\mathrm{p}=3$, and $\gamma_\mathrm{\alpha}=3$ \citep{Gary1993}. This choice represents one-dimensionally adiabatic protons and isothermal electrons. It is consistent with the kinetic slow-mode dispersion relation $\omega_{\mathrm s}\approx k_{\parallel}c_s$ \citep{Gary1993,Narita2015,Verscharen2017}, where  
\begin{equation}
c_s=\sqrt{\frac{3k_\mathrm{B}T_\mathrm{p}+k_\mathrm{B}T_\mathrm{e}}{m_\mathrm{p}}}
\end{equation}
is the phase speed of slow-mode waves in an electron--proton plasma.

\begin{figure}
\plotone{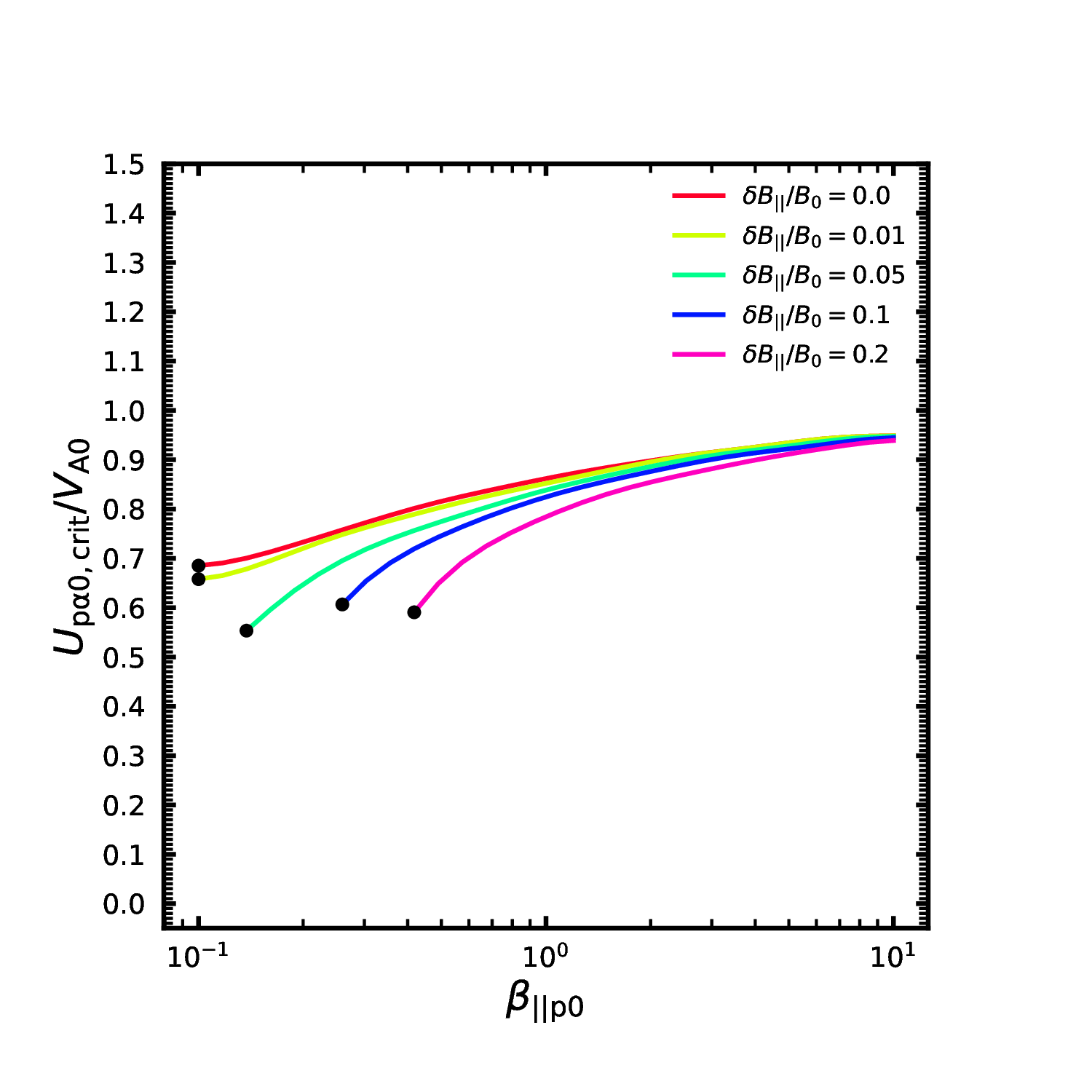}
    \caption{Same as Figure~\ref{fig:fig2} but using $\gamma_\mathrm{e}=1$, $\gamma_\mathrm{p}=3$, and $\gamma_\mathrm{\alpha}=3$.}
    \label{fig:fig5}
\end{figure}

\begin{figure}
    \plotone{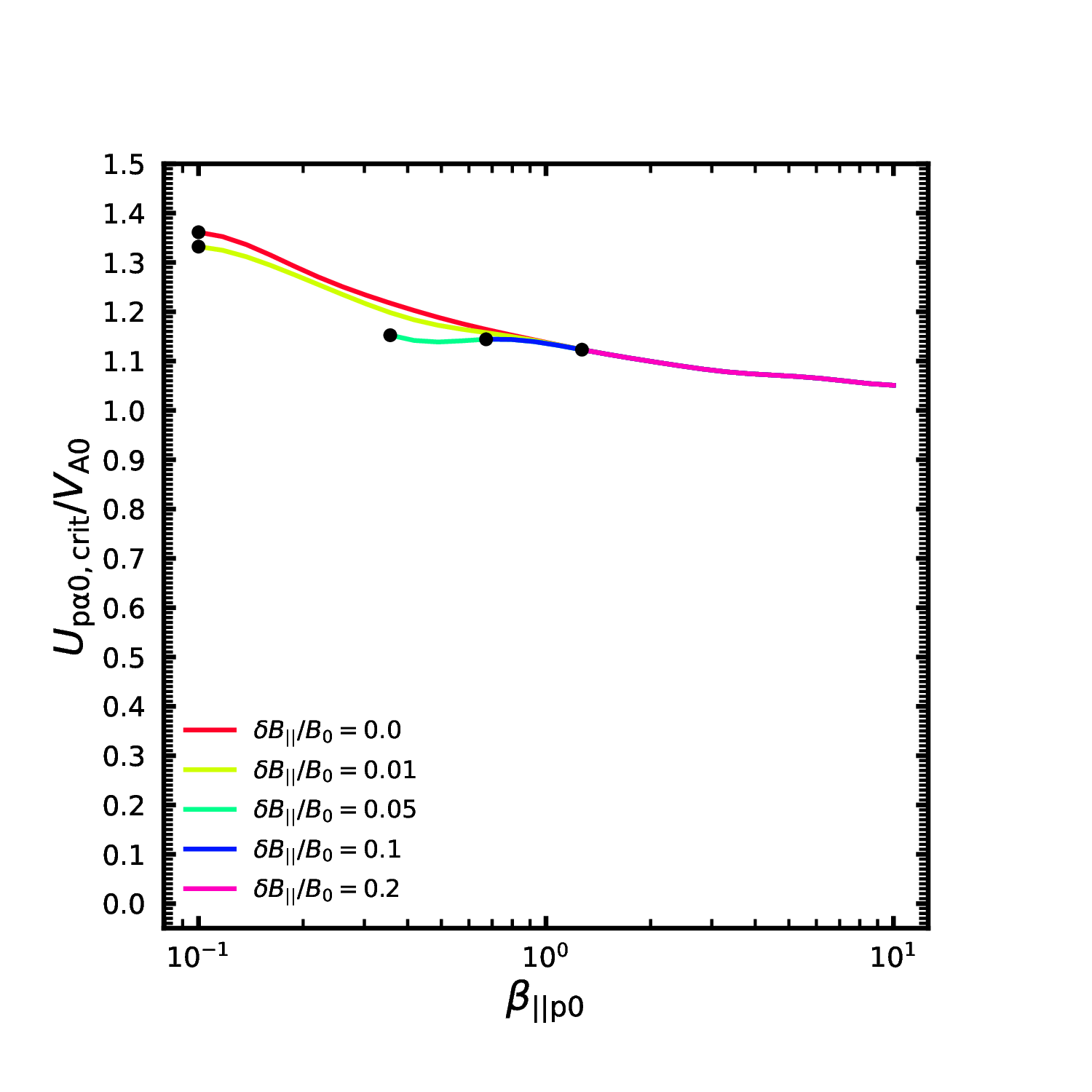}
    \caption{Same as Figure~\ref{fig:fig3} but using $\gamma_\mathrm{e}=1$, $\gamma_\mathrm{p}=3$, and $\gamma_\mathrm{\alpha}=3$.}
    \label{fig:fig6}
\end{figure}

We repeat our analysis with these values for a comparison with the results using polytropic indexes consistent with observations of the effective polytropic indexes (Figures~\ref{fig:fig2} through \ref{fig:fig4}). We show the dependence of $U_{\rm p\alpha 0,crit}/V_{\rm A0}$ on  $\beta_{\rm\parallel p0}$ and $\delta B_{\parallel}/B_0$ for the A/IC and FM/W instabilities in Figures~\ref{fig:fig5} and \ref{fig:fig6}, where we use $\gamma_\mathrm{e}=1$, $\gamma_\mathrm{p}=3$, and $\gamma_\mathrm{\alpha}=3$ .

A comparison between Figures~\ref{fig:fig2} and \ref{fig:fig5} shows insignificant differences in terms of the modified drift thresholds. Only the break-down points extend to lower $\beta_\mathrm{\parallel p0}$ when using the observed polytropic indexes. The differences between Figures~\ref{fig:fig3} and \ref{fig:fig6} are significant in the range of typical $\beta_{\parallel \mathrm p 0}$ values for the solar wind. For $0.5<\beta_{\parallel \mathrm p 0}<1$, the instability  takes place for sub-Alfv\'enic $U_{\mathrm p\alpha 0}$ when $\delta B_{\parallel}/B_0\ge 0.5$, while, under the assumptions applied in Figure~\ref{fig:fig6}, the effective thresholds are always super-Alfv\'enic in this parameter range.

\begin{figure*}
\plotone{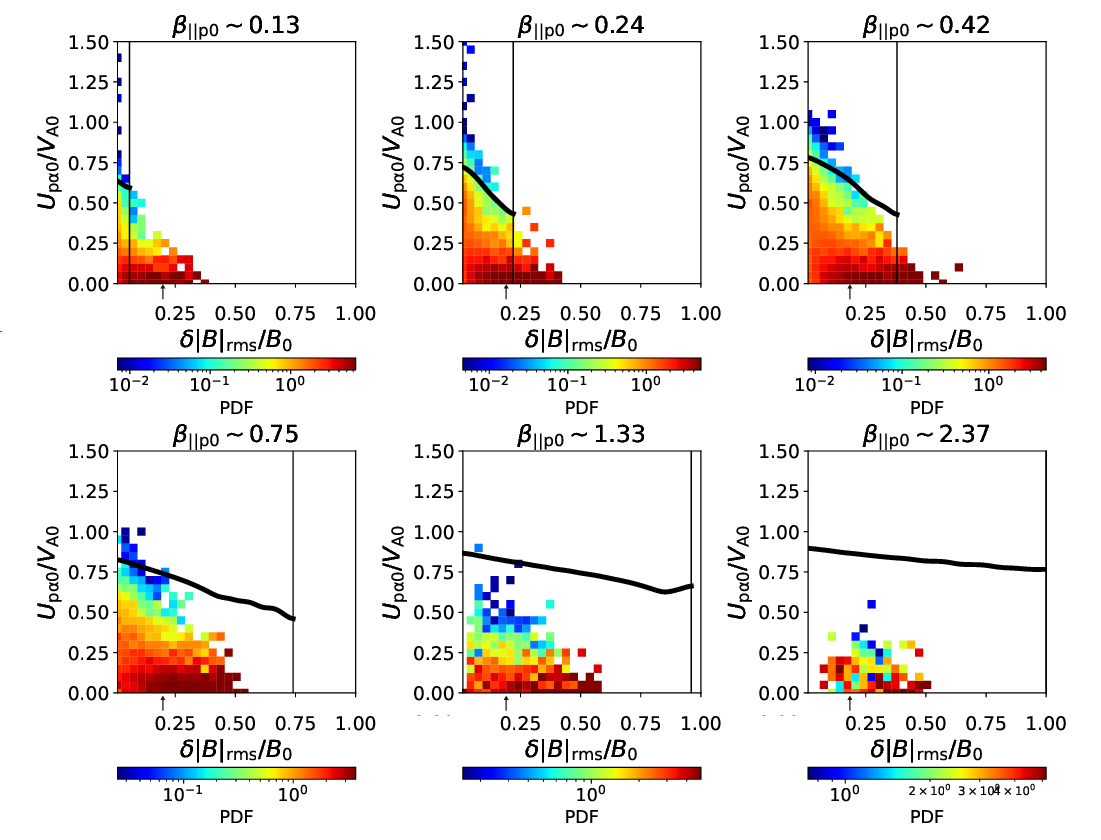}
    \caption{Same as Figure~\ref{fig:fig4} but using $\gamma_\mathrm{e}=1$, $\gamma_\mathrm{p}=3$, and $\gamma_\mathrm{\alpha}=3$. The vertical black lines mark in each panel the amplitude above which our linear model breaks down. The arrows in each panel below the abscissa indicate the amplitude $\delta |\vec B|_{\mathrm{rms}}/B_0=0.2$, the approximate limit of validity of the linear model in all panels.}
    \label{fig:fig7}
\end{figure*}

For completeness, we compare our predictions for $\gamma_\mathrm{e}=1$, $\gamma_\mathrm{p}=3$, and $\gamma_\mathrm{\alpha}=3$  with measurements by the \textit{Wind} spacecraft in Figure~\ref{fig:fig7}. Since the dependence of our results on the particular choice of polytropic indexes is small in the explored parameter space, also this comparison confirms that our theoretical curves limit the data distribution well within the stable parameter space. %The primary difference between Figure~\ref{fig:fig4} and Figure~\ref{fig:fig7} is the extension of the prediction to $\delta\left|{\vec B}\right|_{\rm rms}/B_0\sim 1$ for $\beta_{\rm\parallel p0}<2$.

\begin{figure}
\plotone{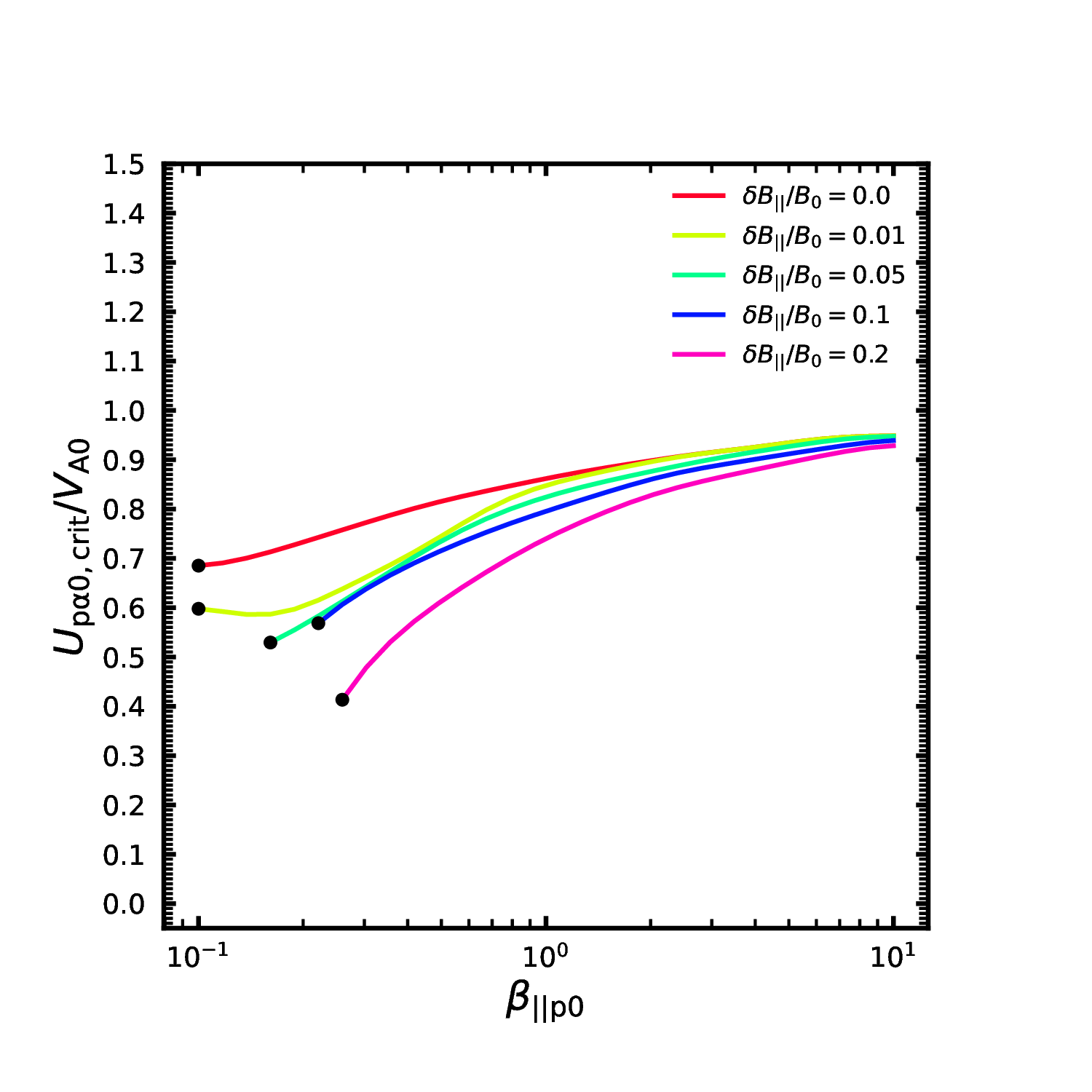}
    \caption{Same as Figure~\ref{fig:fig2} but using $\gamma_\mathrm{e}=1.18$, $\gamma_\mathrm{p}=5/3$, and $\gamma_\mathrm{\alpha}=1.15$.}
    \label{fig:fig8}
\end{figure}

\begin{figure}
\plotone{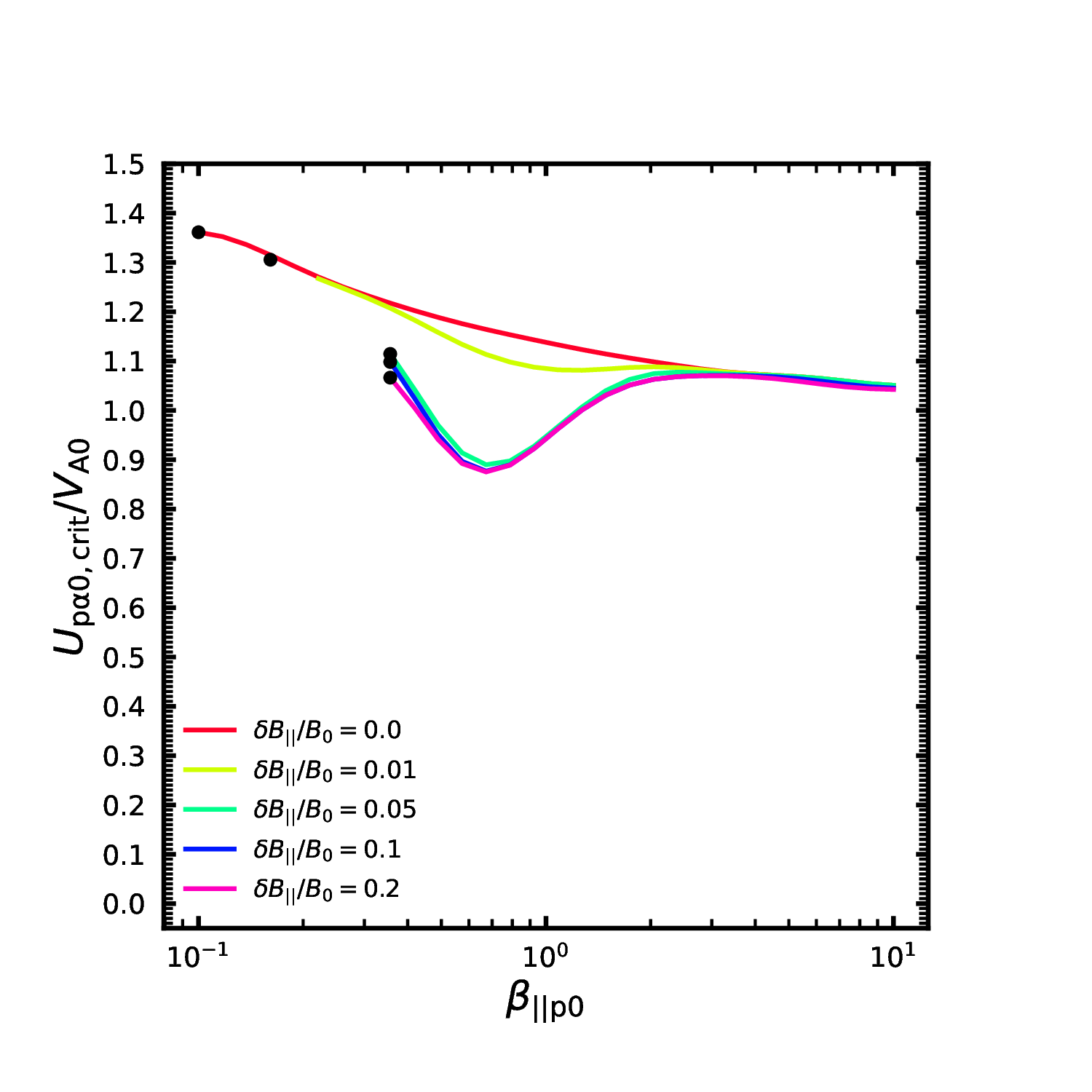}
    \caption{Same as Figure~\ref{fig:fig3} but using $\gamma_\mathrm{e}=1.18$, $\gamma_\mathrm{p}=5/3$, and $\gamma_\mathrm{\alpha}=1.15$.}
    \label{fig:fig9}
\end{figure}

Since the polytropic index of the $\alpha$-particles differs significantly between slow wind and fast wind \citep{Durovcova2019}, Figures~\ref{fig:fig8} and \ref{fig:fig9} show our analysis with the average slow-wind polytropic index $\gamma_\alpha=1.15$ for a comparison with the fast-wind case of $\gamma_\alpha=1.61$ (Figures~\ref{fig:fig2} and \ref{fig:fig3}). There is a noticeable difference between Figure~\ref{fig:fig2} and Figure~\ref{fig:fig8}, in that the smallest effective threshold for $\delta B_{\parallel}/B_0=0.01$ appears at $\beta_{\rm\parallel p0}\sim 0.16$ in the slow-wind case. The differences between Figures~\ref{fig:fig3} and \ref{fig:fig9} are also significant. The effective thresholds decrease to lower values for $\delta B_{\parallel}/B_0\ge 0.05$ in the slow-wind case compared to the fast-wind case.

%% Plasma descriptions typically fall in one of two categories: fluid or kinetic descriptions. These two approaches do not necessarily lead to the same results even at large fluid scales due to the assumed pressure closure of the fluid equations. More specifically for the case of large-scale slow-mode waves, the choice of an appropriate adiabatic index can lead to dispersion relations that are consistent with the kinetic prediction in the large-scale limit. However, both descriptions still predict different  polarization properties of slow-mode waves.We use the term ``kinetic slow-mode wave'' exclusively to denote the predictions from kinetic theory for slow-mode waves at large scales. We note that an alternative convention uses the word ``kinetic'' exclusively for small-scale extensions of the slow-mode wave \citep[e.g.,][]{Narita2015}.
\bibliography{sample631}{}
\bibliographystyle{aasjournal}

%% This command is needed to show the entire author+affiliation list when
%% the collaboration and author truncation commands are used.  It has to
%% go at the end of the manuscript.
%\allauthors

%% Include this line if you are using the \added, \replaced, \deleted
%% commands to see a summary list of all changes at the end of the article.
%\listofchanges

\end{document}